Structural properties, defects and structural phase transition in the ROFeM (R=La, Nd; M=As, P) materials

C. Ma, L.J. Zeng, H.X. Yang, H.L Shi, R.C. Che, C.Y. Liang, Y.B. Qin, G.F. Chen, Z.A. Ren, and J.Q. Li*,

*Beijing National Laboratory for Condensed Matter Physics, Institute of Physics, Chinese Academy of Sciences, Beijing 100080, China*

*Corresponding author. E-mail: LJQ@aphy.iphy.ac.cn*

The structural properties of the ROFeM (R=La, Nd; M=As, P) materials have been analyzed by means of electron diffraction, high-resolution transmission-electron microscopy (TEM) and in-situ cooling TEM observations. The experimental results demonstrate that the layered ROFeM crystals often contain a variety of structural defects, such as stacking faults and small-angle boundaries. The in-situ TEM investigations reveal that, in association with the remarkable spin-density-wave (SDW) instability near 150 K, complex structural transitions can be clearly observed in both crystal symmetry and local microstructure features.

## 1. INTRODUCTION

Since the discovery of copper oxide superconductor in 1986 [1], extensive efforts have been devoted to the search of new high-transition temperature (high-$T_c$) superconducting materials, especially high-$T_c$ systems other than cuprates. The recently discovered superconductors of La[O$_{1-x}$F$_x$]FeP and La[O$_{1-x}$F$_x$]FeAs with the superconducting critical transition temperatures of $T_c$=5K and 26 K respectively [2-6] have attracted quick enthusiasm as a new high-$T_c$ system. Other special merits include surprising effect of F doping on the superconductivity and the magnetic Fe atoms participating into superconductivity. Element substitution, especially the La site replacement by other rare earth elements, gave rise to a variety of notable results in recent investigations. For instance, the superconducting transitions and optimum doping have been extensively investigated in RO$_{1-x}$F$_x$FeAs (R=Ce, Pr, Nd, Sm, Gd) [7-12] and (La$_{1-x}$Sr$_x$)OMAs (M=Fe, Ni) [13,14] materials. Furthermore, without F-doping, Ren et al. have prepared a number of ReFeAsO$_{1-x}$ superconductors under high pressure, and the highest $T_c$ is observed at 55 K in SmFeAsO$_{0.85}$ [15]. These materials therefore are considered as a new family of high $T_c$ superconductors of non-copper-based material with $T_c$ exceeding the upper limit predicted by conventional electron-phonon theory. It is also noted that the (FeAs)-based system contains remarkable competition of ordered ground states along with doping charge carrier [16-19]. The LaOFeAs parent sample shows an anomaly transition at about 150 K in both resistivity and dc magnetic susceptibility measurements [20]. Spin density wave instability, antiferromagnetic super-exchange interactions, spin fluctuation, frustration, and structure phase transitions were discussed based on theoretical calculations and experimental measurements in connection with this notable anomaly [21-23]. In the high-$T_c$ copper oxides, it is generally believed that antiferromagnetism plays a fundamental role in the superconducting mechanism. The parent (non-superconducting) LaOFeAs material is metallic but shows anomalies near 150 K [20]. Neutron scattering measurements demonstrate that LaOFeAs undergoes an abrupt structural distortion below 150 K [21], with the symmetry changing from tetragonal (space group P4/nmm) to monoclinic (space group P112/n) at the low temperature. In this paper, we reported on the investigation of the structural properties of the ROFeM (R=La, Nd; M=As, P) materials prepared under different conditions. In addition, planar defects and structural phase transition were also discussed.

## 2. EXPERIMENTAL METHOD

The polycrystalline samples of ROFeM (R=La, Nd; M=As, P) were prepared by a conventional solid-state-reaction method as reported in previous publications [2,4,6]. X-ray powder diffraction for the structure determination of the as grown materials was performed at room temperature on a Rigaku RINT x-ray diffractometer with Cu $K_\alpha$ radiation. Specimens for TEM observations were prepared by mechanic polishing to a thickness of around 20μm followed by ion milling. In addition, we also prepared some thin samples for electron diffraction experiments simply by crushing the bulk material into fine fragments, which were then supported by a copper grid coated with a thin carbon film. A Tecnai F20 transmission electron microscope, equipped with cooling sample holders, was used for investigating the microstructural properties of these materials from the room temperature down to ~20K.

3. RESULTS

Structural measurements by means of x-ray diffraction and TEM observation have been performed on the samples of the ROFeM (R=La, Nd; M=As, P). The results indicate that all these samples at room temperature have the same tetragonal structure. Figures 1(a), (b) and (c) show the electron diffraction patterns for a LaOFeAs sample, taken along the [001], [100] and [110] zone-axis directions, respectively. All diffraction spots can be well indexed by a tetragonal cell with the lattice parameters of a= 3.962 Å and c= 8.511 Å and a space group of P4/nmm. In Fig. 2(a) and (b), we display the high-resolution images directly showing the atomic structure observed along the $c$-axis and the [100] zone-axis directions, respectively. These images were obtained from the thin regions of the LaOFeAs crystal under the defocus value at around the Scherzer defocus (~-45nm). The atomic positions are therefore recognizable as dark dots. Certain heavy atomic layers along the $c$ direction can be clearly read out in Fig. 2(b). Image simulations, based on the proposed structural model [4], were carried out by varying the crystal thickness from 2 to 5 nm and the defocus value from -30 to -60 nm. A simulated image with the defocus value of -45nm and the thickness of 3nm is superimposed on the image of Fig. 2(b) and appears to be in well agreement with the experimental one. It is hardly surprising that the experimental images have not yielded the perfectly identifiable contrast at the Fe(As) atom positions, owing to the short distance between two adjacent Fe-As atom columns. For instance, the inset of Fig. 2(b) illustrates schematically the projected structure of LaOFeAs crystal along [100] zone-axis direction, where the distance between two adjacent Fe-As atom columns is about 0.13nm which is too short to be resolvable in our electron microscope with a point resolution of ~0.2nm. Hence, FeAs clusters are recognizable as combined dark dots in the experimental image of Fig. 2(b).

Microstructure features of ROFeM (R=La, Nd; As, P) materials depend essentially on the synthesis conditions. Stacking faults as well as some other kinds of planar defects within the a-c plane frequently appear in the crystalline grains. Figure 3(a) shows a bright-field TEM image revealing the presence of numerous stacking faults and grain boundaries in a LaOFeP sample. The planar defects in this kind of materials often locate in the a-b crystal plane and result in apparent structural distortions in the vicinal areas. Figure 3(b) shows a high-resolution TEM image clearly revealing the structural properties of a stacking fault. The structural distortion and lattice mismatch are recognizable nearby this planar defect, though the clear lattice misfit is merely clearly visible within one or two atomic layers. Structural distortion and crystal stress occur in a much large transition area with the size of about 5nm as recognized by the visible contrast anomaly. Another kind of structural defect often appearing in this new superconducting material is the small angle grain boundary. Figure 4(a) shows a high-resolution TEM image illustrating the contrast anomalies and lattice distortions at a grain boundary. Electron diffraction observation shows that a small change of crystal orientations can be clearly seen across this boundary. Figure 4(b) shows the corresponding electron diffraction pattern taken from this typical boundary. The orientation change is estimated to be about 2.4°, and lattice distortion can be clearly recognizable as misfit or bending of crystal layers as indicated by arrows. The frequent appearance of these structural defects originates from the relatively weak bonding along c-axis direction. Structural models for interpreting the arrangements of La, Fe, P, and O atoms nearby these boundaries will be reported in a coming paper.

The parent samples, such as LaOFeAs and NdOFeAs, show a notable transition at about 150 K as demonstrated by the measurements of electric resistivity and dc magnetic susceptibility associated with the spin density wave instability [20]. Previous neutron scattering measurements reveal that LaOFeAs undergoes an abrupt structural transition from tetragonal (space group P4/nmm) to monoclinic (space group P112/n) structure at the low temperature [21]. In order to directly observe the structural changes following with this phase transition, we have performed a series of in-situ TEM investigations on both well-characterized NdOFeAs and LaOFeAs samples. Though we have made numerous attempts to detect the monoclinic distortion within the a-b plane, however, no unambiguous spot splitting in the electron diffraction pattern is visible due to the noticeably small angle alternation from $\beta = 90°$ in the tetragonal phase to $\beta =$ 90.3° in the monoclinic phase [21]. In addition to the neutron diffraction data, we also found certain notable structural changes in both crystal symmetry and microstructure. Figures 5(a) and (b) show the [001] zone-axis electron diffraction patterns taken at 300K and 100K respectively, showing a typical structural change in LaOFeAs observed in many areas. It is remarkable that the Bragg spot at the systematic (100) position, absent at room temperature, become evidently visible at the low-temperature diffraction pattern, this fact suggests a notable alteration in crystal symmetry. Similar structural change was also observed in other ROFeAs samples. In contrast with the previous neutron diffraction data, our low-temperature structural results suggest a monoclinic symmetry with space group of P2 rather than P112/n. On the other hand, our microstructure observations demonstrated that the domain structures arising from the lattice stress in association with the structural phase transition commonly appear in the low-temperature phase. Figure 6 shows two sets of TEM images obtained respectively from the LaOFeAs and NaOFeAs samples, illustrating the presence of complex domain structures at the low-temperature monoclinic phases. These domain boundaries in general are located within the (110) or (100) crystallographic planes. A large fraction of these lamella structures can be understood as twining domains which are energetically preferred for structural relaxation in the present phase transition.

## 4. CONCLUSION

The structural properties of the ROFeM (R=La, Nd, Sm; M=As, P) materials have been analyzed by transmission electron microscopy. It is found that the layered ROFeM crystals contain a variety of common structural defects, including stacking faults and small-angle boundaries. In-situ TEM studies reveal remarkable changes in both crystal symmetry and local microstructure in association with the remarkable spin-density-wave (SDW) instability near 150 K.

Acknowledgments

The authors would like to express many thanks to Prof. Z.X. Zhao and Prof. N.L. Wang for providing samples. This work is supported by the National Science Foundation of China, the Knowledge Innovation Project of the Chinese Academy of Sciences, and the 973 projects of the Ministry of Science and Technology of China.

Figure captions

FIG. 1. Electron diffraction patterns for LaOFeAs taken along the [100] (a), [100] (b) and [110] (c) zone-axis directions, respectively.

FIG. 2. High-resolution TEM images taken respectively along the [001] (a) and [100] (b) zone-axis directions, showing the atomic structural features of LaOFeAs. The inset shows the simulated TEM image. The black rectangle indicates the unit cell. (See text for detail).

FIG.3. (a) Bright field TEM image showing the stacking faults and grain boundaries in a LaOFeP superconductor. (b) High-resolution TEM image taken along the [100] zone axis direction, illustrating the structural distortion for a typical stacking fault in LaOFeP.

FIG. 4. (a) High-resolution TEM image and (b) the corresponding electron diffraction pattern for a small angle grain boundary of $\alpha \cong 2.4°$ for LaOFeP. The white arrows show the dislocation cores at the grain boundary.

FIG. 5. In–situ TEM observations of the parent phase LaOFeAs, taken at 300K (a) and 100K (b) respectively, illustrating the presence of additional (100) spots in the low-temperature monoclinic phase due to a structural phase transition in association with the spin density wave instability.

FIG. 6. In–situ TEM observations on the microstructure changes of the parent phases of NdOFeAs ((a) 300K and (b) 100K) and LaOFeAs ((c) 300K and (d) 100K), demonstrating the remarkable changes of microstructure and structural domains following the phase transition.

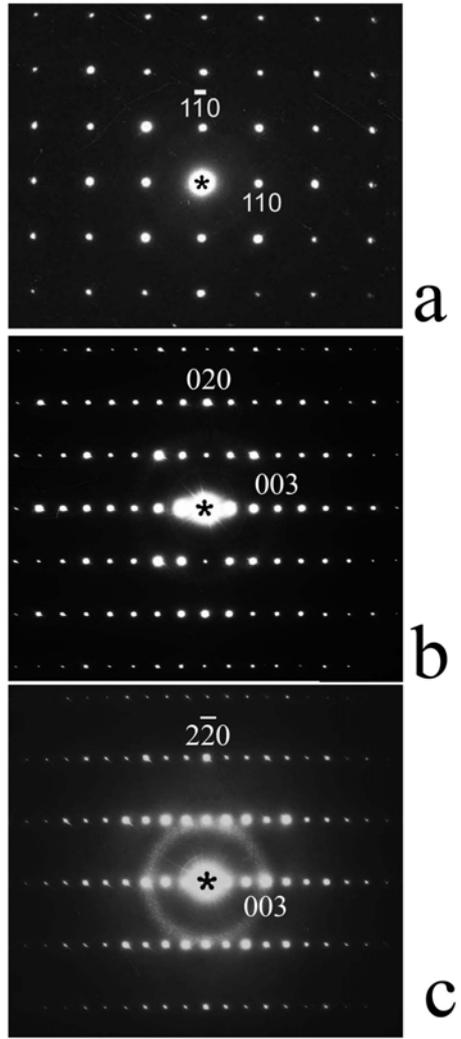

FIG. 1

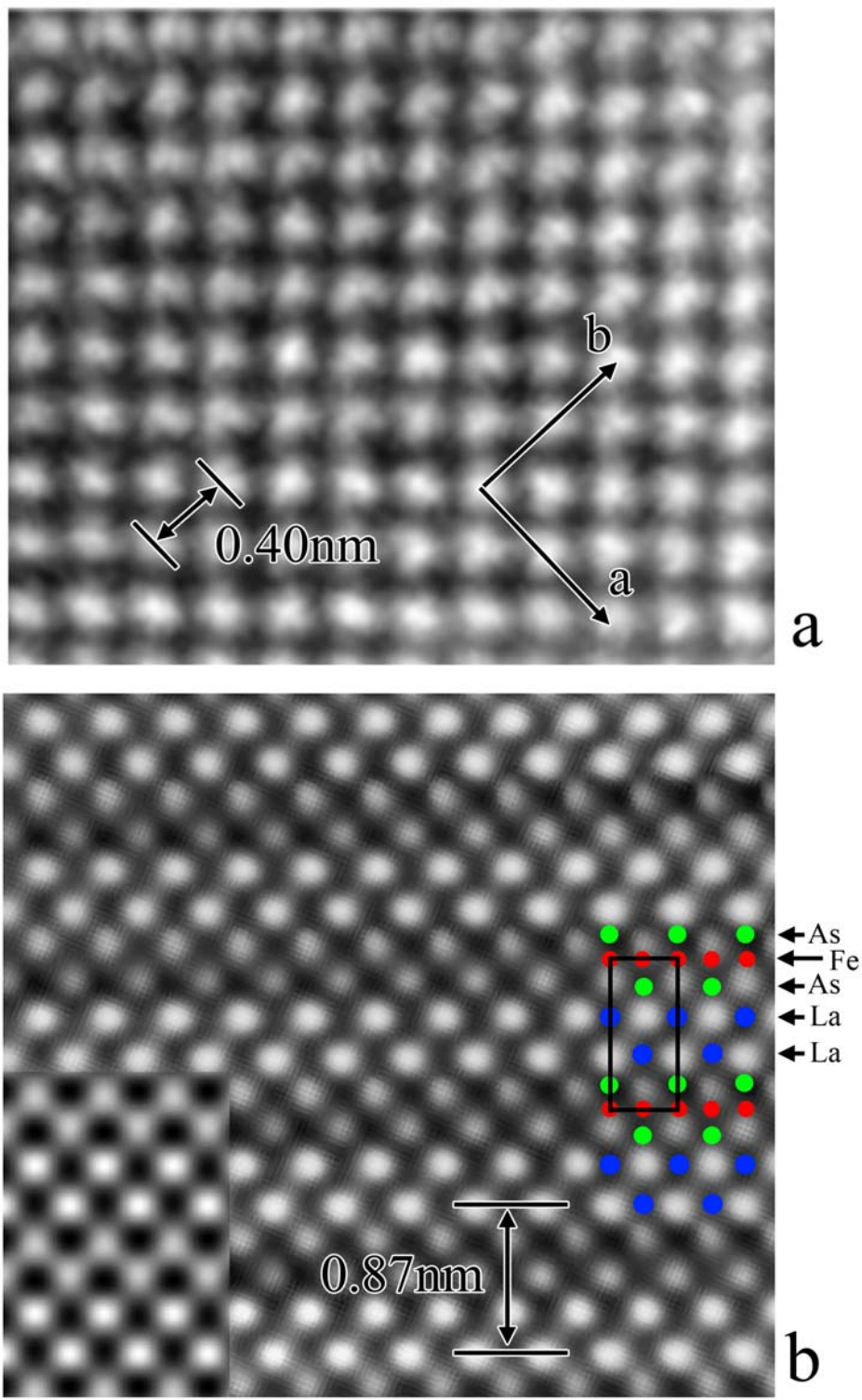

FIG. 2

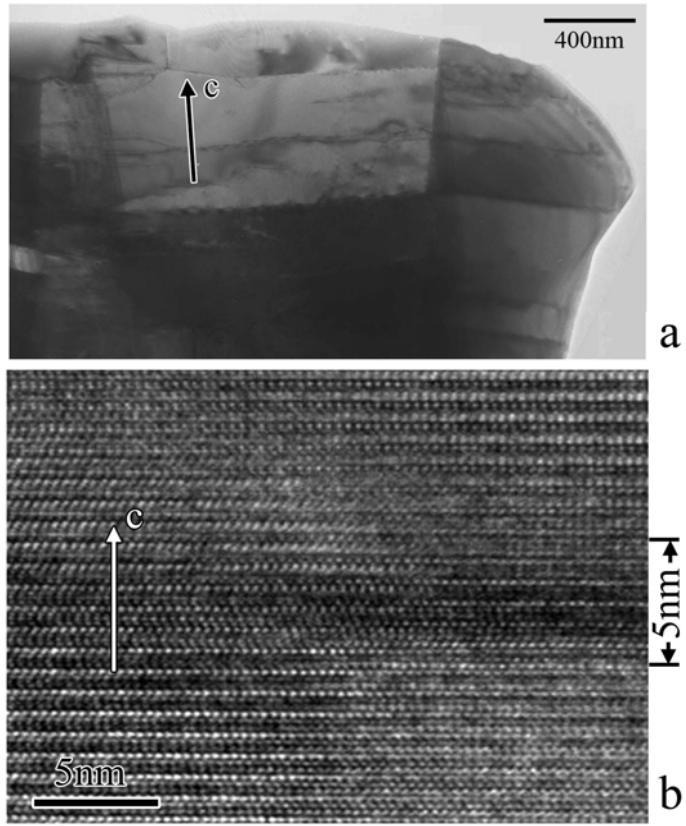

FIG. 3

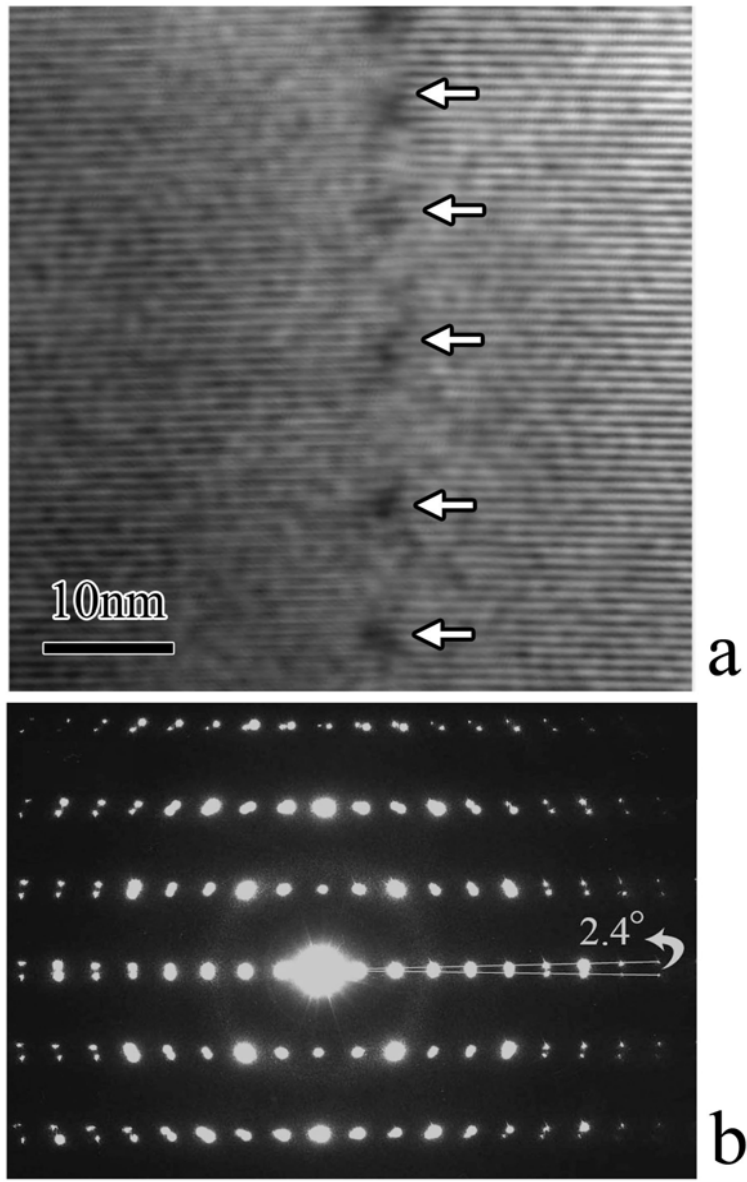

FIG. 4

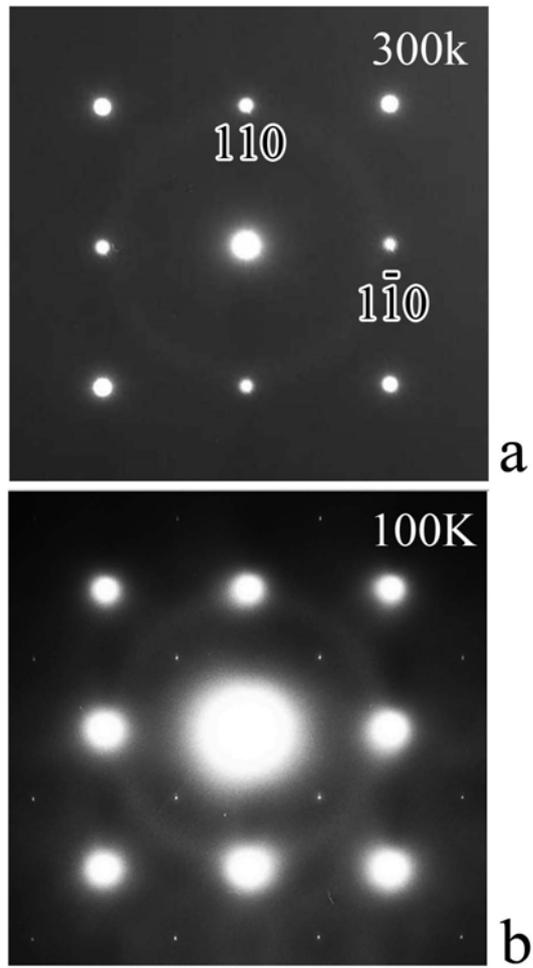

FIG. 5

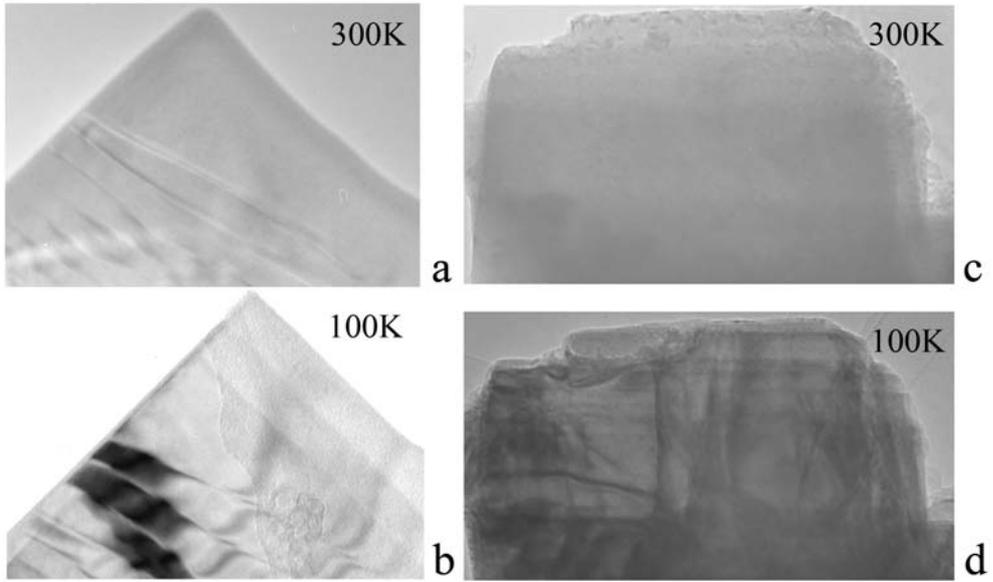

FIG. 6